\documentclass[aps, pra, a4paper,showpacs,twocolumn, 10pt]{revtex4-1}
\usepackage{bbm, amsmath, amssymb, amsthm, bm,textcomp, nicefrac, mathtools,geometry}
\usepackage{amsthm,amssymb,amsmath,graphicx,epsfig,color,verbatim,enumerate,amsthm,appendix}

\newcommand{\ket}[1]{\left|#1\right\rangle}

\newcommand{\bra}[1]{\left\langle #1\right|}

\newcommand{\proj}[1]{\ket{#1}\bra{#1}}

\newcommand{\bea}{\begin{eqnarray}}
\newcommand{\eea}{\end{eqnarray}}

\def\sz{\sigma_z}

\newcommand{\one}{\mbox{$1 \hspace{-1.0mm}  {\bf l}$}}
\def\tr{\mathrm{tr}}

\makeatletter
\newtheorem*{rep@theorem}{\rep@title}
\newcommand{\newreptheorem}[2]{%
\newenvironment{rep#1}[1]{%
 \def\rep@title{#2 \ref{##1}}%
 \begin{rep@theorem}}%
 {\end{rep@theorem}}}
\makeatother

\newreptheorem{theorem}{Theorem}

\begin{document}
\title{Improved quantum metrology using quantum error-correction}
\author{W.\ D\"ur$^1$, M. Skotiniotis$^1$, F. Fr\"owis$^{1,2}$ and B. Kraus$^1$}

\affiliation{$^1$ Institut f\"ur Theoretische Physik, Universit\"at
  Innsbruck, Technikerstr. 25, A-6020 Innsbruck,  Austria\\
  $^2$ Group of Applied Physics, University of Geneva, CH-1211 Geneva 4, Switzerland}
\date{\today}

\begin{abstract}
We consider quantum metrology in noisy environments, where the effect of noise and decoherence limits the achievable gain in precision by quantum entanglement. We show that by using tools from quantum error-correction this limitation can be overcome. This is demonstrated in two scenarios, including a many-body Hamiltonian with single-qubit dephasing or depolarizing noise, and a single-body Hamiltonian with transversal noise.  In both cases we show that Heisenberg scaling, and hence a quadratic improvement over the classical case, can be retained.
Moreover, for the case of frequency estimation we find that the inclusion of error-correction allows, in certain instances, for a finite optimal interrogation time even in the asymptotic limit.
\end{abstract}
\maketitle

{\em Introduction.---}
Parameter estimation is a problem of fundamental importance in physics, with widespread applications in gravitational-wave detectors~\cite{Caves:81, McKenzie:02}, frequency spectroscopy~\cite{Wineland:92, Bollinger:96}, interferometry~\cite{Holland:93, Hwang:02}, and atomic clocks~\cite{Valencia:04, deburgh:05}. Quantum metrology offers a significant advantage over classical approaches, where the usage of quantum entanglement leads to an improved scaling in the achievable precision~\cite{GLM04,Huelga:97}. However, noise and decoherence jeopardize this effect, reducing the quadratic improvement with system size to only a constant gain factor in many scenarios~\cite{Huelga:97,Escher:11,Kolodynski:12,*Kolodynski:13}.

General upper bounds on the possible gain have been derived suggesting that no improvement in the scaling of precision is possible in the presence of uncorrelated, Markovian noise including local depolarizing or dephasing noise~\cite{Escher:11,Kolodynski:12,*Kolodynski:13}. For non-Markovian noise~\cite{Matzusaki:11,*Chin:12}, and noise with a preferred direction transversal to the Hamiltonian evolution~\cite{Chaves:12}, a scaling of ${\cal O}(N^{-3/4})$ and ${\cal O}(N^{-5/6})$ was found respectively, where $N$ denotes the number of probes (see also~\cite{Dorner:12,*Szankowski:12,*Jeske:2013,*Ostermann:13} for results on correlated noise).  This is, however, still below the quadratic improvement attainable in the noiseless case. Moreover, for frequency estimation the optimal interrogation time, i.e. the optimal time to perform the measurement, tends to zero for large $N$ in both these cases making a physical realization for large $N$ impractical.

In this letter we show that, by relaxing the restrictions implicit in standard quantum metrology, namely that the only systems available are the $N$ probes, and the unitary dynamics are generated by local Hamiltonians, the no-go results for the case of uncorrelated, Markovian noise~\cite{Huelga:97,Escher:11,Kolodynski:12,*Kolodynski:13,Chaves:12} can be circumvented, and Heisenberg scaling can be restored.  Specifically, by encoding quantum information into several qubits one can effectively reduce noise arbitrarily at the logical level thereby retaining the Heisenberg limit in achievable precision. The required overhead is only logarithmic, i.e.~each qubit is replaced by $m=\mathcal{O}(\log N)$ qubits. Moreover, we show that in the case of frequency estimation the optimal interrogation time in certain scenarios considered here is finite and independent of the system size, in stark contrast to all frequency estimation protocols studied to date. As the methods we employ can be readily implemented 
experimentally, at least for moderate system
sizes, our result paves the way for the first feasible experimental realization of Heisenberg limited frequency estimation.

\begin{figure}[ht]
\begin{center}
\includegraphics[keepaspectratio, width=8cm]{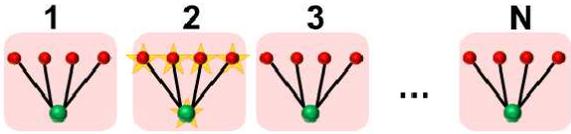}
\end{center}
\caption{Illustration of a quantum metrology scenario using error-correction. We consider $N$ blocks of size $m$ (here $m=5$). In scenario I, all particles in each block are affected by a Hamiltonian $H_I = 1/2 \sigma_z^{\otimes m}$. In scenario II, only the lowest (green) particle of each block is affected by the Hamiltonian $H_{II}=1/2 \sigma_z^{1}$, and $m-1$ ancilla particle (red) are used to generate an effective $m$-body Hamiltonian. In both scenarios, all particles are affected by (local) noise, and each block serves to encode one logical qubit.}
\label{synthesis}
\end{figure}

To be more precise, let us consider a system of $Nm$ qubits which we imagine to be decomposed into $N$ blocks of $m$ qubits with $m$ odd (see Fig.~\ref{synthesis}). First, we consider a class of many-body Hamiltonians, $H_I(m)=1/2\sigma_z^{\otimes m}$, acting on each of the blocks, and uncorrelated, single-qubit dephasing or depolarizing noise (scenario I). Here, and in the following, $\sigma_{x,y,z}$, denote the Pauli operators. We show that, depending on the number of probe systems, $N$, one can choose a sufficiently large $m$ (not exceeding ${\cal O}(\log N)$) such that the Heisenberg limit is achieved even in presence of noise and that the optimal measurement time is constant. Furthermore, we generalize this model to arbitrary local noise and show that for short measurement times the Heisenberg limit can be retrieved. Whereas this model may appear somewhat artificial, it nevertheless serves as a good example to illustrate how quantum error-correction can be used to restore the Heisenberg scaling.

The second, and more physically important, scenario we consider is that of a local Hamiltonian, $H_{II}=1/2\sigma_z^{(1)}$, and local, transversal $\sigma_x$-noise on all qubits.  We show that this scenario can be mapped, for short times, to scenario I, and hence demonstrate how quantum error-correction (and other tools) can be used to arbitrarily suppress noise and restore Heisenberg scaling in precision just as in the noiseless case~\footnote{Note that the reason for obtaining Heisenberg scaling lies in the usage of error-correction, and not in the (logarithmic) increase of system size (which could only lead to a logarithmic improvement). In fact, the Hamiltonians we consider are such that the achievable precession in the noiseless case is independent of $m$.  Moreover, it only depends linearly on $N$, which is also in contrast to the non-linear metrology scheme studied in \cite{Boixo07,*Roy08,*Napolitano11}. We show in both scenarios that we can obtain the same (optimal) precession as in the noiseless 
case.}.
The key idea of our approach lies in the usage of auxiliary particles to encode and protect quantum information against the influence of noise and decoherence as done in quantum error-correction. In addition, the encoding needs to be chosen in such a way that the Hamiltonian acts non-trivially onto the encoded states, such that the information on the unknown parameter is still imprinted onto the system. As long as $H$ is many-body and the noise is local (scenario I), or the Hamiltonian is local and the noise is transversal (scenario II), both conditions can be met simultaneously.

{\em Background.---}We begin by describing the standard scenario in quantum metrology. A probe is prepared in a possibly entangled state of $N$ particles and subsequently undergoes an evolution that depends on some parameter, $\lambda$, after which it is measured.  This process is repeated $\nu$ times and $\lambda$ is estimated from the statistics of the measurement outcomes.
The achievable precision $\delta \lambda$ is lower-bounded by the quantum Cram\'{e}r-Rao bound~\cite{BC94}, $\delta \lambda \geq \frac{1}{\sqrt{\nu {\cal F}(\rho_\lambda)}}$ with ${\cal F}$ the quantum Fisher information (QFI). For local Hamiltonians and uncorrelated (classical) probe states, ${\cal F}=O(N)$ , leading to the so-called standard quantum limit. Entangled probe states, such as the GHZ state, lead to ${\cal F}=O(N^2)$, i.e. a quadratic improvement in precision, the so-called Heisenberg limit. In frequency estimation, time is also a variable that can be optimized, and the quantity of interest in this case is given by ${\cal F}/t$. We refer the reader to Appendix~\ref{Phaseestimation} for details.

In the presence of noise, however, a number of no-go results show that for many uncorrelated noise models, including dephasing and depolarizing noise,
the possible quantum enhancement is limited to a constant factor rather than a different scaling with $N$~\cite{Escher:11, Kolodynski:12,*Kolodynski:13}. To be more specific, we describe the time evolution of the state by a master equation of Lindblad form
\begin{equation}
\dot \rho(t)=-i\lambda[H,\rho]+\sum_{j=1}^{N}{\cal L}_j(\rho),
\label{ME}
\end{equation}
where the action of the single qubit map ${\cal L}_j$ is given by
\begin{equation}
{\cal L}_{j}\rho= \frac{\gamma}{2}(-\rho+\mu_x\sigma_x^{(j)}\rho\sigma_x^{(j)}+\mu_y\sigma_y^{(j)}\rho\sigma_y^{(j)}+\mu_z\sigma_z^{(j)}\rho\sigma_z^{(j)}),
\label{Noise}
\end{equation}
and $\gamma$ denotes the strength of the noise.
The choice $H=H_0=1/2\sum_i \sz^{(i)}$ and $\mu_z=1,\mu_x=\mu_y=0$ corresponds to local unitary evolution and local, uncorrelated, and commuting dephasing noise scenario considered in~\cite{Huelga:97}, whereas for the same Hamiltonian the choice $\mu_x=1,\,\mu_y=\mu_z=0$ corresponds to transversal noise considered in~\cite{Chaves:12}. The choice $\mu_x=\mu_y=\mu_z=1/3$ corresponds to local depolarizing noise. We remark that this approach includes phase estimation for fixed $t=t_0$, and frequency estimation when $t$ can be optimized. 
%When considering a fixed interaction time $t=t_0$, this corresponds to the phase estimation scenario.
%If, however, one has also the possibility to vary, and hence optimize, the interaction time this corresponds to frequency estimation.

For any such scenario investigated so far the attainable precession scales worse than ${\cal O}(N^{-1})$, and the optimal interrogation time tends to zero whenever the noise is not vanishing (see Appendix~\ref{Appfisher} for details).

{\em Quantum metrology with error-correction.---}We now demonstrate that error-correction can be used to recover the Heisenberg limit in the presence of noise in the two scenarios (scenario I and II) mentioned above.
%The relevant parameters to be estimated are the phase or frequency. We show that quantum error-correction (and other tools), can arbitrarily suppress the effects of noise and restore Heisenberg scaling, ${\cal O}(N^{-1})$, in precision achievable in the absence of noise.
For the case of frequency estimation we show that, in certain scenarios, our technique asymptotically allows for a finite, non-zero optimal time to perform measurements in contrast to all current metrological protocols.

{\em Scenario I.---}The evolution of the $Nm$ qubits is governed by the class of Hamiltonians (see Fig.~\ref{synthesis})
%\begin{equation}
$H(m)=\frac{1}{2}\sum_{k=1}^{N} H_k,  H_k=\sigma_z^{\otimes m}$,
%\label{errorH}
%\end{equation}
where $H_k$ acts on block $k$. We assume locality with respect to the blocks, i.e.~this situation is equivalent to having $N$, d-level systems with $d=2^m$. We describe the overall dynamics by Eq.~\eqref{ME}, where the decoherence mechanism is modeled by Eq.~\eqref{Noise}.
In the noiseless case ($\gamma=0$), the maximal attainable QFI is given by ${\cal F}=(\partial\theta/\partial\lambda)^2N^2$ and is obtained by a GHZ-type state, $|GHZ_L\rangle=(|0_L\rangle^{\otimes N} + |1_L\rangle^{\otimes N})/\sqrt{2}$, with $|0_L\rangle = |0\rangle^{\otimes m}$ and $|1_L\rangle = |1\rangle^{\otimes m}$.

Let us now consider the standard metrological scenario in the presence of local dephasing noise, acting on all qubits, where the noise operators commute with the Hamiltonian evolution. In this case Eq.~\eqref{ME} can be solved analytically and the resulting state is given by
%\begin{equation}
%\label{rhot}
$\rho_\lambda(t)={\cal E}_z(p)^{\otimes Nm} \left(U_\lambda|\psi\rangle\langle \psi|U_\lambda^\dagger\right)$,
%\end{equation}
where $U_\lambda=\exp(-i \theta_\lambda H)$ and ${\cal E}_z(p)\rho=p\rho+(1-p)\sigma_z\rho\sigma_z$, with $p=(1+e^{-\gamma t})/2$, are acting on all physical qubits. Phase estimation corresponds to the case where $t=t_0$, for some fixed time $t_0$, and the parameter to be estimated is $\theta_\lambda=\lambda$ resulting from the unitary evolution for time $t_0$. Note that in this case one can start directly with the equation for $\rho_\lambda(t)$, with $p$ being time independent, and a time independent gate $U_\lambda=\exp(-i \lambda H)$ (see Appendix~\ref{Appfisher}). As the subsequent discussion is independent of whether $p$ is time dependent or not, we simply write $p$ in the following whenever it does not lead to any confusion.

We now encode each logical qubit in $m$ physical qubits. On each block of $m$ qubits we make use of an error-correction code, similar to the repetition code, capable of correcting up to $(m-1)/2$ phase-flip errors (recall that we chose $m$ to be odd), with code words
%\begin{align}
$\ket{0_L} = (|0_x\rangle^{\otimes m} + |1_x\rangle^{\otimes m})/\sqrt{2}$,
$|1_L\rangle = (|0_x\rangle^{\otimes m} - |1_x\rangle^{\otimes m})/\sqrt{2}$,
where $|0_x\rangle=(|0\rangle + |1\rangle)/\sqrt{2}, |1_x\rangle=(|0\rangle - |1\rangle)/\sqrt{2}$.
The error-correction procedure consists of projecting onto subspaces, $P_{\vec k}$, spanned by $\{\sigma_z^{\vec{k}}|0_x\rangle^{\otimes N}, \sigma_z^{\vec{k}}|1_x\rangle^{\otimes N}\}$, where ${\vec k}=(k_1,\ldots,k_m)$ with $k_i\in \{0,1\}$. Here, $\sigma_z^{\vec{k}}$ denotes the $m$ qubit local operator, $\sigma_z^{k_1}\otimes \sigma_z^{k_2}\ldots \otimes \sigma_z^{k_m}$. After obtaining outcome ${\vec k}$ the correction operation $\sigma_z^{\vec{k}}$ is applied. As long as fewer than $(m-1)/2$ $\sigma_z$ errors occur we obtain no error at the logical level. Otherwise, a logical $\sigma_z^{(L)}$ error occurs. Hence, the noise at the logical level can again be described as logical phase-flip noise, ${\cal E}_z^{(L)}(p)\left(\rho\right)=p_L\rho+(1-p_L)\sigma_z^{(L)}\rho\sigma_z^{(L)}$, with
\begin{equation}
\label{pL}
p_L=\sum_{k=0}^{\frac{m-1}{2} } \binom{m}{k}p^{m-k} (1-p)^{k},
\end{equation}
where $p_L>p$ for $p>1/2$. %, i.e.~noise at the logical level is suppressed.
For small errors, i.e.~$(1-p) \ll 1$,  the Taylor expansion of $p_L$ can be approximated by
%\bea
%\label{Eq:Plapprox}
$p_L=1- \binom{m}{\frac{m+1}{2}}(1-p)^{\frac{m+1}{2}}+{\cal O}[(1-p)^{\frac{m}{2}+1}]$,
%\eea
to leading order in $(1-p)$.  That is, noise at the logical level is exponentially suppressed.

We now consider a logical GHZ state, $|GHZ_L\rangle=(|0_L\rangle^{\otimes N} + |1_L\rangle^{\otimes N})/\sqrt{2}$, as input state~\footnote{See also Refs.~\cite{Frowis:11,*Frowis:12} for studies on the stability of this state under noise.}. At the logical level, $H_k$ acts as a logical $\sigma_z^{(L)}$ operation, $H_k|0_L\rangle=|0_L\rangle$, $H_k|1_L\rangle=-|1_L\rangle$, and the (time) evolved state, $|\psi_\lambda^L\rangle=U_\lambda |GHZ_L\rangle=(e^{-iN\theta_\lambda/2}|0_L\rangle^{\otimes N} + e^{iN\theta_\lambda /2}|1_L\rangle^{\otimes N})/\sqrt{2}$, remains within the logical subspace. The state is then subjected to phase noise acting on each of the qubits. After correcting errors within each block of $m$ qubits, phase noise at the logical level is reduced (see above). The state after error-correction is given by $\rho_\lambda^L=[{\cal E}_z^L(p_L)]^{\otimes N}\left(\proj{\psi_\lambda^L}\right)$. As a result, the situation is equivalent to the standard phase estimation scenario
with
a single-qubit, $\sigma_z$ Hamiltonian and local phase noise, where the error probability is, however,
exponentially suppressed.

Let us now bound the precision for both phase and frequency estimation. As $\rho_\lambda$ is of rank 2 the Fisher information can be easily calculated~\cite{Kolodynski:12,*Kolodynski:13} (see Appendix~\ref{Appfisher}), and for phase estimation one finds
%\begin{equation}
${\cal F}(\rho_\lambda)=(2p_L-1)^{2N} N^2$.
%\label{QFIphase}
%\end{equation}
In contrast to the standard scenario, where the strength of the noise is independent of $N$, here $p_L$ can be made arbitrarily close to 1. Hence, one encounters a quadratic scaling and thus recovers the Heisenberg limit. For any fixed value of $p$ and $m$, we have Heisenberg scaling up to a certain, finite-system size, $N_{\rm max}$. For example, for $p=1-10^{-3}$ we find $(2p_L-1) = 1- \epsilon_L$ with $\epsilon_L \approx 6 \times 10^{-6}, 2 \times 10^{-8}, 1.3 \times 10^{-15}$ for $m=3,5,11$ respectively. Hence, $(2p_L-1)^{2N} = {\cal O}(1)$,
i.e. a constant close to 1,
as long as $2N \epsilon_L \ll 1$. Thus, for $N$ up to $N_{\rm max} = {\cal O}(1/\epsilon_L)$ our error-correction technique would yield Heisenberg scaling in precision.
More importantly, if $m=\mathcal{O}(\log N)$,  and using the approximation $\binom{m}{\frac{m+1}{2}}<2^m$, it can be shown that $(2p_L-1)^{2N} \rightarrow 1$ and ${\cal F} \approx N^2$ for $N\to \infty$ as long as $4N(2\sqrt{1-p})^m \ll 1$. Thus, the QFI can be stabilized, and the Heisenberg limit is attained, with only a logarithmic overhead ~\footnote{By logarithmic overhead we mean that each particle is replaced by $m=\mathcal{O}(\log N)$ particles. Note that in practice there is no need for a separate error-correction step followed by measurements to determine the parameter, but a single measurement with proper re-interpretation suffices.}.

If instead of phase estimation we consider frequency estimation, i.e.~$\theta_\lambda=\lambda t$, we obtain (see Appendix~\ref{Appfisher})
%\begin{equation}
${\cal F}(\rho_\lambda)=t^2 (2p_L(t)-1)^{2N}N^2$,
%\label{QFIfreq}
%\end{equation}
where $2p_L(t)-1=e^{-\gamma_L(m,\gamma,t)t}$, and $\gamma_L(m,\gamma,t)$ is the noise parameter at the logical level. Assuming that $\gamma t\ll1$
the optimization of $\mathcal{F}/t$ over $t$ can be easily performed. Assuming that $m=\mathcal{O}(\log N)$ the optimal interrogation time and the bound on precision for an arbitrary number of $m$ are presented in Appendix~\ref{Appfisher} . We find that the optimal interrogation time decreases for larger system sizes $N$. However, $t_{\rm opt}$ gets larger with increasing $m$, and can hence be much more feasible in practice. Assuming that $\gamma t \ll 1$ and $m=\mathcal{O}(\log N)$, $p_L$ can be approximated using Stirling's formula and we find
%\begin{equation}
$t_{\mathrm{opt}}= \frac{N^{-\frac{2}{m}}}{2\gamma m^{\frac{2}{m}}} \to \frac{1}{2\gamma e^2}$.
%\label{topt}
%\end{equation}
Thus the optimal measurement in our scenario can be performed at a finite time for large $N$. This is to be contrasted with the optimal times for previously considered frequency estimation scenarios, based on GHZ and other entangled states, where $t_{\rm opt} \to 0$ for large $N$~\cite{Huelga:97,Chaves:12}.
The maximum QFI per unit time is then given by
%\bea \label{QFImax}
$\left(\frac{\mathcal{F}}{t}\right)_{\mathrm{opt}}=\frac{N^{2(1-\frac{1}{m})}}{2\gamma m^{\frac{2}{m}}} \to \frac{N^2}{2 \gamma e^2}$,
%\eea
and the Heisenberg limit is approached for $N\to \infty$.

In Appendix~\ref{AppA} we show that any kind of local error can be treated in this way. This is done by using an error-correction code that corrects for arbitrary single-qubit errors rather than just bit-flip errors, and where the Hamiltonian still acts as a logical $\sigma_z^{(L)}$ operator on the codewords. We find that one obtains Heisenberg scaling for short measurement times,  $t\propto N^{-1/2}$.

{\em Scenario II.---}Let us now consider the physically more relevant scenario where the Hamiltonian is given by $H=H_0=1/2\sum_i \sigma_z^{i}$, and transversal noise~\footnote{We remark that in practical situations, parallel noise will often be dominant. The optimal measurement is typically transversal to the Hamiltonian, and imperfections in the measurement lead to parallel noise.}.
%As mentioned before, it was recently shown that the QFI scales as ${\cal O}(N^{5/3})$ and that this scaling can be achieved using the GHZ state~\cite{Chaves:12}.

We now show that the Heisenberg limit is attainable also in this case.  To this aim, we attach to each of the system qubits $m-1$ ancilla qubits, not affected by the Hamiltonian, that may also be subjected to (directed) local noise (see Fig.~\ref{synthesis}). In practice, this may be achieved using qubits associated with different degrees of freedom (e.g. other levels in an atom), or another type of physical system. The situation is hence similar to scenario I, i.e.~we have $Nm$ qubits that are decomposed into $N$ blocks of size $m$. The Hamiltonian is given by
%\begin{equation}
$H=\frac{1}{2}\sum_{k=1}^{N} H_{k},\quad H_k=\sigma_z^{(1)}\otimes I^{\otimes m-1}$.
%\label{H2}
%\end{equation}
and we consider transversal noise acting on each of the $Nm$ qubits, see Eqs.~(\ref{ME},\ref{Noise}).
%, with $\mu_x=1$ and $\mu_y=\mu_z=0$, and acts on each of the $Nm$ qubits. The overall evolution is given by Eq.~\eqref{ME}.

In the following we show that the above situation can indeed by mapped precisely to the situation considered in scenario I. To this end, imagine that after preparing the entangled (encoded) resource state (i.e.~a logical GHZ state $|GHZ_L\rangle$), we apply an entangling unitary operation ${\cal U}^\dagger$ to all qubits, allow them to freely evolve according to Eq.~\eqref{ME}, and apply ${\cal U}$ before the final measurement. The result is that the evolution takes place with respect to a unitarily transformed master equation
%\begin{equation}
$\dot \rho=-i\lambda [\tilde H,\rho]+\sum_{j=1}^{Nm}\tilde{\cal L}_j(\rho)$,
%\label{unitaryME}
%\end{equation}
where $\tilde H={\cal U}H{\cal U}^\dagger$, and $\tilde{\cal L}_{j}\rho=\frac{\gamma}{2}\left(-\rho+({\tilde {\cal U}}\sigma_x^{(j)}{\tilde {\cal U}}^\dagger)\rho({\tilde {\cal U}}\sigma_x^{(j)}{\tilde {\cal U}}^\dagger)\right)$.
Here, ${\cal U}=\otimes_{k=1}^N {\cal V}_k$ with ${\cal V}_k=\prod_{j=2}^{m}CX^{(1,j)}$, where ${\cal V}_k$ acts on a single block, and $CX=({\rm Had}\otimes {\rm Had})CP({\rm Had}\otimes {\rm Had})^\dagger$ with $CP={\rm diag}(1,1,1,-1)$ the controlled phase gate, and ${\rm Had}$ the Hadamard operation. The action of such a transformation has been studied and applied in the context of simulating many-body Hamiltonians~\cite{Dur:08}. It is straightforward to verify that~\cite{Dur:08}
%\begin{align}\nonumber
${\cal U}H_k{\cal U}^\dagger={\cal V}_k H_k{\cal V}_k^\dagger = \sigma_z^{(1)}\otimes \sigma_x^{\otimes m-1}$,
${\cal U}\sigma_x^{(j)}{\cal U}^\dagger={\cal V}_k\sigma_x^{(j)}{\cal V}_k^\dagger = \sigma_x^{(j)}$,
%\end{align}
where the transformed Hamiltonian, ${\cal U}H_k{\cal U}^\dagger$, acts within a block. Up to Hadamard operations on particles $2,\ldots m$, this corresponds to the situation described in scenario I, i.e.~an $m$-qubit Hamiltonian, $H_k=\sigma_z^{\otimes m}$, and local, single-qubit noise ($X$ noise on particle 1 and $Z$ noise on all ancilla particles). As shown in Appendix~\ref{AppA} one can achieve Heisenberg scaling for any local noise model using logical GHZ states as input states. This implies that we also achieve Heisenberg scaling---at least for short measurement times, $t\propto N^{-1/2}$~\footnote{If $\delta t^2 N \not\ll 1$, then we have to take higher order terms in the solution of the master equation into account. This leads to parallel noise of ${\cal O}(\delta t^2)$ and limits the maximal $N$ until which Heisenberg scaling can be achieved~\cite{Chaves:12}.}---for transversal local noise, where the required block size is again $m= {\cal O}(\log N)$.

{\em Experimental realization.---}
 We now consider a simplified version of scenario II, where only particles that are affected by the Hamiltonian are affected by noise, i.e.~noise is part of the coupling process, involving a two-qubit error correction code which can be easily  demonstrated experimentally.
%This is the case, for example, when noise is part of the coupling process.
The error correction code with $|0_L\rangle = |0\rangle|0_x\rangle$, $|1_L\rangle = |0\rangle|1_x\rangle$ as codewords, is capable of correcting arbitrary $\sigma_x$ errors occurring on the first qubit, while the Hamiltonian still acts as a logical $\sigma_z^{L}$ after the transformation ${\cal U}$. This opens the way for simple proof-of-principle experiments in various set-ups, including trapped ions or photonic systems, where a total of $2N$ qubits prepared in a GHZ-type states suffices to obtain a precision $O(N^{-1})$.

%In this case, one can use the simple two-qubit error correction code, $|0_L\rangle = |0_x\rangle|0\rangle$, $|1_L\rangle = |0_x\rangle|1\rangle$, to correct for arbitrary $\sigma_x$ errors occurring on the first qubit, while the Hamiltonian still acts as a logical $\sigma_z^{L}$ after the transformation ${\cal U}$. This opens the way for simple proof-of-principle experiments in various set-ups, including trapped ions or photonic systems, where a total of $2N$ qubits prepared in a GHZ-type states suffices to obtain a precision $O(N^{-1})$.

{\em Conclusion and outlook.---}We have demonstrated that quantum error-correction can be applied in the context of quantum metrology and allows one to restore Heisenberg scaling in several scenarios. This includes the estimation of the strength of a multi-qubit Hamiltonian in the presence of arbitrary independent local noise, as well as a single-body Hamiltonian in the presence of transversal noise. In the latter case, an improvement in the precision from ${\cal O}(N^{-5/6})$, previously shown in~\cite{Chaves:12}, to ${\cal O}(N^{-1})$ is demonstrated. Furthermore, for frequency estimation we have shown that the interrogation time can be finite and independent of $N$ in contrast to all previously known parameter estimation protocols.
This demonstrates that, even though recent general bounds suggest a limitation of the possible gain in noisy quantum metrology to a constant factor for dephasing or depolarizing noise, this is actually not the case in general. It remains an open question whether tools from quantum error-correction can also be applied in other metrology scenarios, most importantly in the context of estimating local Hamiltonians in the presence of parallel (phase) or depolarizing noise~\footnote{Note that our results from scenario II can not be directly applied in the case of parallel or depolarizing noise. Using a unitary transformation to obtain a many-body Hamiltonian also transforms parallel noise to correlated noise, that can not be corrected by the error-correction code used here.}.

\textit{Acknowledgements.---}This work was supported by the Austrian Science Fund (FWF): P24273-N16, Y535-N16, SFB F40-FoQus F4012-N16, J3462.

\textit{Note added.---}After completing this work we learned about independent work using similar approaches~\cite{Arad:13,Kessler:13,Ozeri:13}.

\section*{Appendices}
In the following appendices we provide detailed calculations for the main results in the paper.  Specifically, Sec.~\ref{Phaseestimation} includes a brief review of phase and frequency estimation. In Sec.~\ref{Appfisher} we discuss the quantum Fisher information (QFI), and provide a proof of finite, non-zero optimal time and Heisenberg scaling in precision %(Eqs.~(11,12) in the letter)
for scenario I.  In Sec.~\ref{AppA} we show how our error-correcting scheme is capable of dealing with arbitrary local noise provided we consider short measurement times.
\appendix
\section{Phase and frequency estimation}
\label{Phaseestimation}
We start by describing the standard scenario in quantum metrology. A probe is prepared in a possibly entangled state of $N$ particles. It undergoes an evolution that depends on some parameter, $\lambda$, and the probe is measured afterwards. The overall process is repeated $\nu$ times and $\lambda$ is estimated from the statistics of the measurement outcomes.
The achievable precision in the estimation of $\lambda$, $\delta \lambda$, which measures the statistical deviation of the estimator from the actual parameter, is lower-bounded by the quantum Cram\'{e}r-Rao bound~\cite{BC94},
\begin{equation}
\label{QCR}
\delta \lambda \geq \frac{1}{\sqrt{\nu {\cal F}(\rho_\lambda)}},
\end{equation}
where ${\cal F}$ denotes the quantum Fisher information of the state $\rho_\lambda$ resulting from the evolution of the initial state of the $N$ probes~\cite{BC94}. Note that the bound can be reached asymptotically, i.e.~for $\nu\rightarrow \infty$.

In the noiseless case we have $\rho_\lambda=U_\lambda \rho_0 U_\lambda^\dagger$, where $U_\lambda =e^{-i\theta_\lambda H}$ for some Hamiltonian $H$. In the literature one distinguishes between phase estimation, where $\theta_\lambda=\lambda$ is the parameter to be estimated, and frequency estimation, where $\theta_\lambda=\lambda t$ and the frequency $\lambda$ has to be estimated. In the later case not only the number of particles, $N$, counts as a resource but the additional resource of the total running time, $T=\nu t$, has to be taken into account. The QFI for pure input states, $\rho=|\psi\rangle\langle \psi|$, is then given by ${\cal F}(\rho_\lambda)= (\frac{\partial\theta_\lambda}{\partial\lambda})^2 4 {\rm Var}(H)_{\rho_\lambda}$,
where ${\rm Var}(H)_\rho$ denotes the variance of $H$ with respect to the state $\rho$. If the aim is to estimate frequency the bound in precision, Eq.~\eqref{QCR}, can be written as $\delta \lambda \sqrt{T} \geq \frac{1}{\sqrt{{\cal F}(\rho_\lambda(t))/t}}$ in order to account for the total running time $T$. Here, the QFI obtained per unit time, ${\cal F}(\rho_\lambda(t))/t$, has to be optimized over time leading to an optimal interrogation time $t_{\rm opt}$.

\section{Fisher Information}
\label{Appfisher}
In this section we briefly recall the definition and some properties of the quantum Fisher information, ${\cal F}(\rho)$. The latter is defined as~\cite{BC94}
\bea {\cal F}(\rho)=\tr(\rho^\prime L_\rho)=\tr(\rho L_\rho^2),\eea
where the Hermitian operator $L_\rho$ is the symmetric logarithmic derivative of $\rho$ and is defined via the equation
\bea \frac{d\rho}{d\lambda}=\rho^\prime\equiv \frac{1}{2}(\rho L_\rho +L_\rho \rho).\eea
Writing $\rho$ in its spectral decomposition as $\rho=\sum_i p_i \proj{\Psi_i}$, it can be easily seen that \bea L_\rho=2\sum_{j,k:p_j+p_k\neq 0}\frac{1}{p_j+p_k}\bra{\Psi_j}\rho^\prime \ket{\Psi_k}\ket{\Psi_j}\bra{\Psi_k},\eea
which leads to
\bea {\cal F}(\rho)=2\sum_{j,k:p_j+p_k\neq 0}\frac{1}{p_j+p_k}|\bra{\Psi_j}\rho^\prime \ket{\Psi_k}|^2.\eea

The computation of the QFI is in general hard since the diagonalization of $\rho$ is required. However, there exist several upper bounds on the Fisher information in the literature~\cite{Escher:11,Kolodynski:12,*Kolodynski:13}.

Throughout the paper we consider the situation where $\rho_\lambda=U_\lambda {\cal E}(\rho_0) U_\lambda^\dagger$, with $U_\lambda =e^{-i \lambda H}$ for some Hamiltonian, $H$, and initial state, $\rho_0$. Here, ${\cal E}$ denotes a completely positive, trace-preserving map that is independent of the parameter to be estimated. Such a map could result, for example, from solving the master equation, in case the unitary and dissipative evolution are commuting, from approximating the solution of the master equation for short times, or from a time-independent evolution which the system is subject to.

In the case of phase estimation, i.e.~$\theta_\lambda =\lambda$, $\rho_\lambda^\prime =-i[H,\rho_\lambda]$ and one obtains for the QFI
\bea
\label{AppFishPhase}
{\cal F}(\rho_\lambda)=2\sum_{j,k:p_j+p_k\neq 0}\frac{(p_j-p_k)^2}{p_j+p_k}|\bra{\Psi_j}H \ket{\Psi_k}|^2.
\eea
For frequency estimation, where $\rho_\lambda^\prime =-it^2 [H,\rho_\lambda]$, one obtains
\bea
\label{AppFishFre}
{\cal F}(\rho_\lambda)=2t^2 \sum_{j,k:p_j+p_k\neq 0}\frac{(p_j-p_k)^2}{p_j+p_k}|\bra{\Psi_j}H \ket{\Psi_k}|^2.
\eea
Note that the sums in Eqs.~(\ref{AppFishPhase},\ref{AppFishFre}) run over ${\cal O}(2^N)$ terms. Furthermore, if $\rho_\lambda=(\sigma_\lambda)^{\otimes N}$, for some single qubit state, $\sigma_\lambda$,
(which is the case for local Hamiltonians and local noise acting on a product state as input state)
it can be shown that $F[(\sigma_\lambda)^{\otimes N}]=N F[(\sigma_\lambda)]$, and the Fisher information scales linearly in $N$.

In the noiseless case, where $\rho_\lambda=U_\lambda (\rho_0) U_\lambda^\dagger$, it can easily be seen that for pure input states Eqs.~(\ref{AppFishPhase},\ref{AppFishFre}) reduce to
\bea {\cal F}(\rho_\lambda)=4 {\rm Var}(H)_{\rho_\lambda} \\  {\cal F}(\rho_\lambda)=t^2 4 {\rm Var}(H)_{\rho_\lambda}\eea respectively, where ${\rm Var}(H)_{\rho}=\langle H^2 \rangle_{\rho}-\langle H \rangle_{\rho}^2$ denotes the variance of $H$ with respect to the state $\rho=|\psi\rangle\langle \psi|$.

It follows that for uncorrelated (classical) input states, the precision of phase and frequency estimation is bounded by $\delta \lambda \geq \frac{1}{\sqrt{\nu N}}$ and $\delta \lambda \sqrt{T}\geq \frac{1}{\sqrt{ N}}$ respectively, as the QFI can only scale as ${\cal O}(N)$ for such states. This is also known as the standard quantum limit. In contrast, a scaling of ${\cal O}(N^2)$ for the QFI is possible for entangled probe states, leading to the so-called Heisenberg limit with an attainable precision of $\delta \lambda = 1/(\sqrt{\nu}N)$ and $\delta \lambda \sqrt{T}\geq \frac{1}{N}$ respectively. The bound is achieved by preparing the probe in the Greenberger-Horne-Zeilinger (GHZ) state, $\ket{GHZ}=(\ket{0}^{\otimes N}+\ket{1}^{\otimes N})/\sqrt{2}$.

When taking noise into account, Heisenberg scaling can however no longer be achieved. For instance, as shown in~\cite{Escher:11}, if we consider noise described by Eq. (2) in the main text, where $\gamma\neq 0$ and $\mu_z=1,\mu_x=\mu_y=0$, the ultimate precision in frequency estimation is given by $\delta\lambda \sqrt{T} \geq \sqrt{\frac{2\gamma}{N}}$.
In contrast the best classical strategy yields a bound $\delta\lambda \sqrt{T}\geq\sqrt{2\gamma e/N}$, i.e. only a gain by a constant factor is found. Notice that the GHZ state---which is optimal in the noiseless case---has an optimal interrogation time $t_{\mathrm{opt}}=\frac{1}{2N\gamma}$, but does not provide any gain in the noisy case.
For the case of transversal noise the achievable precision and corresponding interrogation time were shown to be $\delta\lambda \sqrt{T}\geq\sqrt{\frac{(9\gamma)^{1/3}}{2N^{5/3}}}$, and $t_{\rm opt}=(3/\gamma N)^{1/3}$ respectively~\cite{Chaves:12}. Note that in both cases, the interrogation time tends to zero as $N$ gets large, making a physical realization of the optimal protocol very challenging. In fact, for larger measurement times it has been shown that the scaling ${\cal O}(N^{-5/6})$ cannot be achieved~\cite{Chaves:12}.

We now compute the QFI, in the case of phase estimation, for scenario I where $\rho_\lambda={\cal E}_z(p)^{\otimes N}(U_\lambda \proj{GHZ} U_\lambda^\dagger)$.
The only two non-vanishing eigenvalues of $\rho_\lambda$ are
\bea
p_{0,1}=\frac{1}{2}(1\pm (2p-1)^{2N}),
\eea
and the corresponding eigenstates are $\ket{\Psi_{0,1}}=e^{-i\theta_\lambda N/2}\ket{0}^{\otimes N}\pm e^{i\theta_\lambda N/2} \ket{1}^{\otimes N}$. All other eigenvalues are zero and do not contribute to the QFI. This can be seen by considering the kernel of $\rho_\lambda$ which is given by the span of $\left\{\ket{\vec{k}}|\,|\vec{k}|\neq 0,N\right\}$. As $ \bra{\Psi_{0,1}}H\ket{\vec{k}}=0$ for $|\vec{k}|\neq0, N$ and
$ |\bra{\Psi_{0}}H\ket{\Psi_{1}}|=N/2$, we obtain for the QFI
\begin{equation*}
{\cal F}(\rho_\lambda)=4 (p_0-p_1)^2|\bra{\Psi_0}H \ket{\Psi_1}|^2=(2p-1)^{2N} N^2.
\end{equation*}
Similarly, for frequency estimation we have
\begin{equation*}
{\cal F}(\rho_\lambda)=t^2(2p(t)-1)^{2N} N^2.
\end{equation*}

We now consider the process at the logical level, i.e.~where error-correction has been employed and we obtain $p(t)=p_L(t)$ with $p_L(t)$ given by Eq.~(3) in the main text. The optimal interrogation time and QFI can be straightforwardly determined. Using the approximation $p_L=1- \binom{m}{\frac{m+1}{2}}(1-p)^{\frac{m+1}{2}}+{\cal O}[(1-p)^{\frac{m}{2}+1}]$ as indicated in the main text), together with ${\cal F}(\rho_\lambda)=(2p_L-1)^{2N} N^2$, and assuming that $m=\mathcal{O}(\log N)$ and $\gamma t$ is small, optimization of  $\mathcal{F}/t$ over $t$ yields for the optimal interrogation time and precision bound:
\bea
t_{\mathrm{opt}}&=&\left(\frac{1}{2\binom{m}{\frac{m+1}{2}}\left(\frac{\gamma}{2}\right)^{\frac{m+1}{2}}(3+Nm)}\right)^{\frac{2}{m+2}}\\
\left(\frac{\mathcal{F}}{t}\right)_{\mathrm{opt}}&=&\frac{N^2}{\left(2\binom{m}{\frac{m+1}{2}}\left(\frac{\gamma}{2}\right)^{\frac{m+1}{2}}(3+Nm)\right)^{\frac{2}{m+2}}}\left(\frac{Nm+2}{Nm+3}\right)^{2N}.    \nonumber
\eea
Using Stirling's approximation we obtain $t_{\mathrm{opt}}= \frac{N^{-\frac{2}{m}}}{2\gamma m^{\frac{2}{m}}} \to \frac{1}{2\gamma e^2}$ and $\left(\frac{\mathcal{F}}{t}\right)_{\mathrm{opt}}=\frac{N^{2(1-\frac{1}{m})}}{2\gamma m^{\frac{2}{m}}} \to \frac{N^2}{2 \gamma e^2}$ as stated in the main text. Notice that above equations are only valid for sufficiently large $m$, $m=\mathcal{O}(\log N)$, and we have used $m=\ln N$ to arrive at the final result. 

As a second example let us compute the QFI for the standard metrology scenario with a local Hamiltonian, $H=\sum_i\sigma^{(i)}_z$, and depolarizing noise described by $p=e^{-2\gamma_L \delta t/3}$ (see Sec.~\ref{AppA}).  As in this case the local noise commutes with the local Hamiltonian we have $\rho_\lambda= U_\lambda^{\otimes N} [D(p)^{\otimes N} (\rho_0)] (U^\dagger_\lambda)^{\otimes N}$. If the initial state, $\rho_0$, is the GHZ state the eigenbasis, $\{\ket{\Psi_i}\}$, of $[D(p)^{\otimes N} (\rho_0)]$ is given by $\ket{\vec{k}}$, where $|\vec{k}|\neq 0,N$, and the two states $\ket{\Psi_{0,1}}=1/\sqrt{2}(\ket{0}^{\otimes N}\pm  \ket{1}^{\otimes N})$.
This can be easily verified as
\bea
D(p)^{\otimes N} (\rho_0)=\left(\frac{1-p}{2}\right)^N \rho_0+&&\nonumber \\
\sum_{k=0}^{N-1} p^k \left(\frac{1-p}{2}\right)^{N-k} \sum_P P [\one\otimes \one \ldots \otimes  \tr_{1,\ldots N-k}(\rho_0)]P,&&\nonumber
\eea
where the sum runs over all possible permutations, and $\tr_{1,\ldots N-k}(\rho_0)$ denotes the reduced state of qubits $(N-k+1),\ldots, N$.
Thus, the eigenstates of $\rho_\lambda$ are the states $U_\lambda \ket{\Psi_i}$. Since $U_\lambda$ commutes with $H$, we need to determine the overlaps $\bra{\Psi_i}H \ket{\Psi_j}$. As $H$ is diagonal in the computation basis this overlap vanishes for $i\neq j$ unless $\{i,j\}$=$\{0,1\}$. Thus, the QFI is given by
\begin{equation}
{\cal F}=4 \frac{(p_0-p_1)^2}{p_0+p_1}|\bra{\Psi_0}H \ket{\Psi_1}|^2=\frac{p^{2N}}{(\frac{1+p}{2})^N + (\frac{1-p}{2})^N} N^2,
\label{AppdepolQFI}
\end{equation}
where $p_{0,1}=\frac{1}{2}\left[\left(\frac{1+p}{2}\right)^N+\left(\frac{1-p}{2}\right)^N \pm p^N\right]$ denote the eigenvalues of $\ket{\Psi_{0,1}}$ respectively.

\section{Local noise}
\label{AppA}

Here we show that the error-correction method presented in scenario I, with $H=H(m)$ given by $H(m)=\frac{1}{2}\sum_{k=1}^{N} H_k,  H_k=\sigma_z^{\otimes m}$, apply to any kind of local noise if we consider short measurement times.
We first consider local depolarizing noise, and then demonstrate that the results also hold for arbitrary local noise. Depolarizing noise is described by the completely positive map
\bea
{\cal E}(\rho)=p\rho+\frac{(1-p)}{4}\sum_{i=0}^3 \sigma_i \rho \sigma_i=p\rho+\frac{(1-p)}{2}\one.
\eea

On each block, one uses an error-correction code corresponding to graph states~\cite{Hein:06}, e.g.~a 5-qubit code corresponding to a ring graph, that can correct an arbitrary error on one qubit~\cite{Gottesman:97,Grassl:02,Schlingeman:01,Schlingeman:02}. Using such a code in a concatenated fashion allows one to reduce noise at the logical level to an arbitrary degree as long as $\gamma<\gamma_{\rm Code}$. In fact, one finds that the noise at the logical level is logical depolarizing noise~\cite{Kesting:13}. Let $|G\rangle$ be a graph state, $|G\rangle = \prod_{(j,k)\in E}U_{jk}|+\rangle^{\otimes m}$, where $U_{jk}={\rm diag}(1,1,1,-1)$ is a phase gate acting on qubits $j,k$, and the graph is described by edges $(j,k) \in E$. Defining the logical states
\begin{align}
|0_L\rangle &= (|G\rangle + \sigma_z^{\otimes m}|G\rangle)/\sqrt{2},\nonumber\\
|1_L\rangle &= (|G\rangle - \sigma_z^{\otimes m}|G\rangle)/\sqrt{2},
\end{align}
the action of $H_k$ on these logical states is given by $H_k|0_L\rangle=|0_L\rangle$ and $H_k|1_L\rangle=-|1_L\rangle$. That is $H_k$ acts as a logical phase flip, $\sigma_z^{(L)}$.
If we only consider the noisy part of the evolution, which on each block is given by $\sum_{k=1}^m{\cal L}_k$, this leads to depolarizing noise acting on each of the qubits,
$\tilde\rho_t=[{\cal D}(p)]^{\otimes N}(\rho)$ with
\bea
{\cal D}(p)\rho=p\rho+\frac{1-p}{4}\sum_{j=0}^3\sigma_j^{(k)}\rho\sigma_j^{(k)}=p\rho+\frac{1-p}{2} \one
\eea
and $p=e^{-2\gamma t/3}$.

As the noise and unitary evolution do not commute the master equation can not be easily solved as in the case of dephasing noise. However, we might approximate the solution for short evolution times using the Trotter expansion. For times $\delta t^2 N \ll 1$
the output state is well approximated by
\begin{equation}
\label{Apprhot}
\rho(\delta t)=[{\cal D}(p)]^{\otimes N} \left(U_{\delta t}|\psi\rangle\langle \psi|U_{\delta t}^\dagger\right).
\end{equation}
If we apply error-correction before performing the final measurement, the noise for each block acts as depolarizing noise at the logical level with parameter $p_L > p$ for $p$ sufficiently large~\cite{Kesting:13}. That is, the situation at the logical level is equivalent to a standard metrology scenario with local Hamiltonian, $\sigma_z$, and depolarizing noise described by $p_L=e^{-2\gamma_L \delta t/3}$. The QFI in this case is given by (see Sec.~\ref{Appfisher})
\begin{equation}
{\cal F}=\frac{p_L^{2N}}{(\frac{1+p_L}{2})^N + (\frac{1-p_L}{2})^N} N^2,
\label{depolQFI}
\end{equation}
and can be approximated, for $p_L$ sufficiently close to 1, as ${\cal F} \approx p_L^{3N/2} N^2$. Note that this QFI would be obtained whenever the state $\rho_\lambda$ is described by Eq.~\eqref{Apprhot}.

Noise at the logical level can be exponentially reduced when using a concatenated error-correction code~\cite{Hein:05,Kesting:13}. For the concatenated 5-qubit code with $n$ concatenation levels the block size is $m=5^n$. For $n=1$ one finds that the probability, $q$, to have no error at the logical level is well approximated by~\cite{Hein:05,Kesting:13}
\bea
\label{qL}
q_L=q^5+5q^4(1-q),
\eea
where $q=(1+3p)/4$, and $q_L=(1+3p_L)/4$ for depolarizing noise. That is all events that correspond to zero error (probability $q^5$) or one error at one of the qubits (5 instances, each with probability $q^4(1-q)$) can be corrected by the code leading to no error at the logical level. A simple concatenation of Eq.~\eqref{qL}
leads to the logical error probability when using a concatenated code~\cite{MikeIke}. One finds that the effective noise parameter, $\gamma_L$, is exponentially suppressed~\cite{Hein:05}.
Similar to dephasing noise, for $m={\cal O}(\log N)$ we again recover a quadratic scaling of the QFI and hence of the achievable precision.

A generalization to arbitrary local noise is straightforward. The reason is that quantum error-correction codes can deal with any kind of local noise, as long as the probability $q$ for no error is sufficiently large. In fact, as shown in~\cite{Kesting:13}, Pauli noise acting on the individual qubits is mapped to (logical) Pauli noise at the logical level. The probability to have no error at the logical level is given by Eq.~\eqref{qL} and the above approximations still hold when dealing with concatenated codes.
Alternatively, one can actually bring arbitrary local noise process described by a completely positive map, or noise in a master equation described by a local Liovillian, to a standard form corresponding to local depolarizing noise. This is done by means of depolarization, i.e.~by applying certain local unitary operations randomly, and might increase the noise level by a constant factor~\cite{Dur:05}.
\bibliographystyle{apsrev4-1}
\bibliography{parestprl}
\end{document}